\documentclass[debug]{rmaa}

\title{Time variation of the O/H radial gradient
in the galactic disk based on planetary nebulae} 

\author{
  W. J. Maciel,\altaffilmark{1} 
  and R. D. D. Costa,\altaffilmark{1}}

\altaffiltext{1}{Instituto de Astronomia, Geof\'\i sica e Ci\^encias
                 Atmosf\'ericas, Universidade de S\~ao Paulo, Brazil.}

\shortauthor{Maciel \& Costa}
\shorttitle{Time variation of the O/H radial gradient}

\fulladdresses{
\item W. J. Maciel, and R. D. D. Costa: 
    Instituto de Astronomia, Geof\'\i sica e Ci\^encias
    Atmosf\'ericas, Universidade de S\~ao Paulo - 
    Rua do Mat\~ao 1226, CEP 05508-090, S\~ao Paulo SP, Brazil
    (maciel@astro.iag.usp.br, roberto@astro.iag.usp.br.)}

\listofauthors{W. J. Maciel & R. D. D. Costa}
\indexauthor{Maciel, W. J.}
\indexauthor{Costa, R. D. D.}

\abstract{
The controversy on the time variation of the radial abundance gradients can in principle 
be settled by estimating the gradients from planetary nebulae (PN) ejected by central 
stars (CSPN) with different ages. In this work, we consider four  samples of CSPN whose 
lifetimes have been estimated using three different methods and estimate the oxygen 
abundance gradients for these objects. The results suggest some small differences between 
the younger and older CSPN. The younger objects have similar or slightly higher oxygen 
abundances compared with the older objects, and the gradients of both groups are similar 
within the uncertainties. Therefore, the O/H radial gradient has not changed appreciably 
during the lifetime of the objects considered, so that PN gradients are not expected to 
be very different from the gradients observed in younger objects, which seems to be 
supported by recent observational data.
}

\resumen{ }

\addkeyword{ISM: planetary nebulae: general}
\addkeyword{Galaxies: Abundances}
\addkeyword{Galaxies: Disk}
\addkeyword{ISM: abundances}

\begin{document}
\maketitle

\section{Introduction}
\label{section1}

Radial abundance gradients are observed in the galactic disk based on abundance measurements
of several chemical elements in a variaty of astronomical objects (Henry \& Worthey 
\citeyear{henry1999}, Maciel \& Costa \citeyear{mc2010}, Maciel et al. \citeyear{mrc2012}). 
The main chemical elements are oxygen, neon, sulphur and argon in photoionized nebulae, and 
iron in stars. However, recent work also includes data on many  other elements, especially in
cepheids, such as Ba (Andrievsky et al. \citeyear{andrievsky2013}), several $\alpha$-elements, 
iron-peak elements, and even heavier elements (see for example Cescutti et al. 
\citeyear{cescutti2007}). In view of the variety of chemical elements and objects, the 
gradients are especially important as constraints of chemical evolution models, which is 
stressed by the fact that the gradients do not appear to be constant, but present both space 
and time variations, which increases the number of constraints that must be satisfied by 
realistic models.

The problem of the time variation of the abundance gradients - here understood as the 
{\it slope} of the gradients, usually measured in dex/kpc - is particularly important, as 
it is instrumental in distinguishing between different chemical evolution models. For example,
the model by Chiappini et al. (\citeyear{chiappini2001}) predicts a continuous steepening 
of the gradients with time, while models by Hou et al. (\citeyear{hou2000}) predict just 
the opposite behaviour. Estimating the gradients at different  epochs is a difficult problem, 
as it implies some knowledge of the ages of the objects involved, apart from their chemical 
abundances and distances. As is well known, stellar ages are uncertain, especially considering 
evolved objects, with ages greater than about 2 to 3 Gyr (see for example Soderblom 
\citeyear{soderblom2010}, \citeyear{soderblom2009}).

While the present day gradient may be determined on the basis of the observed abundances
of young objects, such as HII regions, with typical ages of a few million years, or cepheid 
variables, with ages up to a few hundred million years (see for example Maciel et al.
\citeyear{mlc2005}), the gradient at past epochs is more appropriately studied on the basis 
of planetary nebulae (PN) and open clusters. PN are formed by progenitor stars with masses 
in the approximate range of 0.8 to 8 $M_\odot$ on the main sequence, so that their ages would 
be expected to vary from about 1 Gyr to several Gyr, as indicated for example by Table~7 of 
Stasi\'nska (2004). On the other hand, open clusters have an even broader time range from a few 
million years up to several Gyr (see for example Andreuzzi et al. \citeyear{andreuzzi2011} 
or the most recent version of the open cluster catalogue by Dias et al. \citeyear{wilton}). 

In a previous work (Maciel et al. \citeyear{mcu2003}), we have studied 
the time variation of the abundance gradients using  planetary nebulae  based largely 
on their classification according to the Peimbert scheme (Peimbert 
\citeyear{peimbert1978}). While this scheme succeeds in predicting average ages and central 
star masses (see for example Stasi\'nska \citeyear{stasinska2004}, Table 7), the derived 
results are of difficult interpretation, since the adopted classification implies some 
ambiguity in the stellar properties for many objects.
  
In order to improve this investigation, we have developed five methods to estimate
individual ages of CSPN, as opposed to the average ages implied by the Peimbert
types. These methods are based either on the measured nebular abundances or on the kinematic 
properties of the nebulae and their central stars. In the first paper (Maciel, Costa \& Idiart 
\citeyear{mci2010}), we developed three methods based on the chemical abundances, 
using (1) an age-metallicity-distance  relation, (2) a simpler  age-metallicity relation,  
and  (3)  a relation between the central star masses and the nebular nitrogen abundances
(see also Maciel et al. \citeyear{mcu2003}, \citeyear{mlc2005}). More recently (Maciel, 
Rodrigues \& Costa \citeyear{mrc2011}), we developed two kinematic methods based on the 
velocity dispersion-age relation from the Geneva-Copenhagen survey (Holmberg et al.
\citeyear{holmberg2009}) using (4) the galactic rotation curve or (5) the $U$, $V$, $W$ 
velocity components. 
 
The five Methods~1--5 above were applied to different samples of galactic PN with known properties, 
and the age distributions were determined in each case.  Based on these results, most CSPN 
in the galactic disk have ages under 6 Gyr, and the age distribution has a prominent peak,
but its exact location depends on the adopted method. In the present work, we 
selected the most accurate of the five methods developed so far, namely Methods~1, 3, and 5, 
and applied them to different samples of galactic PN in order to investigate the time variation 
of the radial abundance gradients. Method~2 fails to produce a prominent peak in contrast
with the most reliable methods, and Method~4 depends on several assumptions regarding the
actual PN rotation curve, so that these methods are not included in the present work.
In section~2 we summarize the main characteristics of the methods considered in this investigation 
and in section~3 we describe the PN samples adopted. In section~4 we describe the procedure used 
to estimate the gradients at different times and the main results are presented and discussed,
and in section~5 the main conclusions are stated.

\section{Age determination of CSPN}
\label{section2}

\subsection{Method 1: The Age-metallicity-radius relation}

Method 1 was discussed in detail by Maciel et al. (\citeyear{mcu2003}), where it was also
called Method~1, and applied to a sample of planetary nebulae in the galactic disk by Maciel et 
al. (\citeyear{mci2010}). It uses a relationship between the ages of the PN progenitor stars, the 
nebular abundances, and the galactocentric distances developed by Edvardsson et al. 
(\citeyear{edvardsson}). The original relation involves the [Fe/H] metallicities, which are obtained 
from the oxygen abundances measured in the nebulae using a relation developed by Maciel et al. 
(\citeyear{mcu2003}). The obtained age distribution as applied to the original sample of 234 
nebulae by Maciel et al. (\citeyear{mci2010}) is shown in Figure~1. It can be seen that a 
well-defined peak exists in the age distribution at about 4-5 Gyr.

   \begin{figure}
   \centering
   \includegraphics[angle=-90, width=9.0cm]{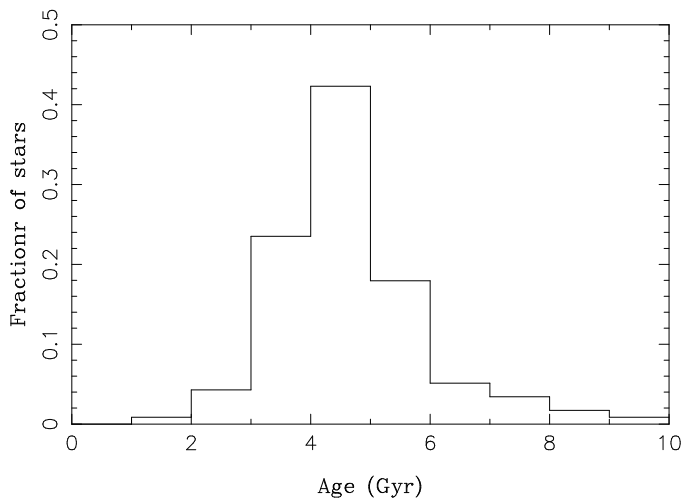}
   \caption{Age distribution of the central stars of planetary nebulae 
      for Method 1, based on an age-metallicity-galactocentric distance relation.}
   \label{fig1}
   \end{figure}

\subsection{Method 3: The N/O $\times$ CSPN mass relation}

Method 3 is based on a relationship between the mass of the planetary nebula central star
and the N/O abundance ratio measured in the nebulae also developed by Maciel et al. 
(\citeyear{mcu2003}), and based on an earlier analysis of a selected sample of galactic
nebulae (Cazetta and Maciel \citeyear{cazetta2000}). The method also assumes an initial mass-final 
mass relation for the central stars, and we adopt here Case B of Maciel et al. (\citeyear{mci2010}), 
which uses the mass-age relation by Bahcall \& Piran (\citeyear{bahcall}),  and was considered the
most realistic case compared with their Case~A, which assumes a much simpler mass-age relation
leading to very large lifetimes for the more massive stars.

Figure~2 shows the derived age distribution for this method as applied by Maciel et al.
(\citeyear{mci2010}) to a sample of 122 galactic nebulae. The distribution is similar to the previous 
case in the sense that most objects have ages lower than about 6 Gyr, and there is a prominent peak,
but the average ages are lower than in the case of Method~1, which reflects the fact
that in Method~3 the lifetimes of the massive stars are lower and the probability of finding stars 
at larger lifetimes is smaller.

   \begin{figure}
   \centering
   \includegraphics[angle=-90, width=9.0cm]{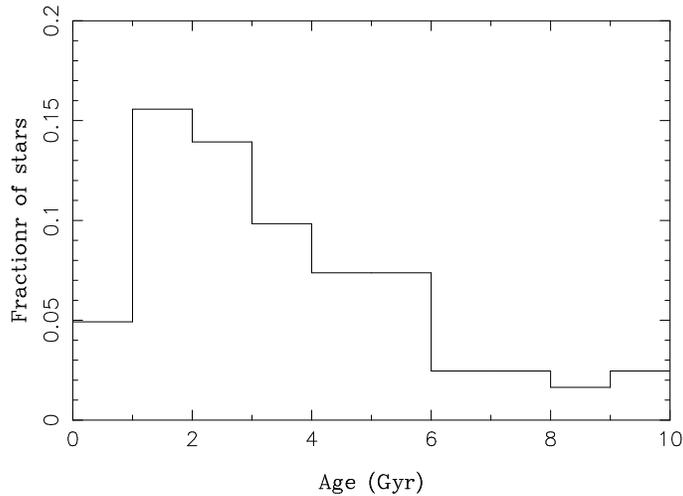}
   \caption{Age distribution of the central stars of planetary nebulae 
      for Method 3, based on an relation between the central star
      mass and the nebular N/O ratio.}
   \label{fig2}
   \end{figure}

\subsection{Method 5: The U,V,W, velocity components}

Method~5 corresponds to Method~2 of Maciel et al. (\citeyear{mrc2011}) and is a kinematic
method, in which we have determined the  $U$, $V$, $W$, velocity components and the 
total velocity $T$, as well as the velocity dispersions $\sigma_U$, $\sigma_V$, $\sigma_W$, and 
$\sigma_T$. Accurate relations between the velocity dispersions and the stellar
ages were obtained by the Geneva-Copenhagen Survey of the Solar Neighbourhood (cf. Nordstr\"om et al. 
\citeyear{nordstrom}, Holmberg et al. \citeyear{holmberg2007}, \citeyear{holmberg2009}),
which allowed the determination of the ages of the CSPN in our sample. 

Figure~3 shows the age distribution for Method~5 applied to a large sample of planetary
nebulae with measured radial velocities from the catalogue by Durand et al. (\citeyear{durand})
which contains 867 objects. In this case we have adopted the total $T$ velocities, and 
distances from the  distance scale by Stanghellini et al. (\citeyear{ssv2008}, SSV), which are
available for 403 objects in the catalogue of Durand et al. (\citeyear{durand}). Again, the
distribution shows a prominent peak, although it is somewaht displaced towards
lower ages compared to the distributions shown in the previous figures.

   \begin{figure}
   \centering
   \includegraphics[angle=-90, width=9.0cm]{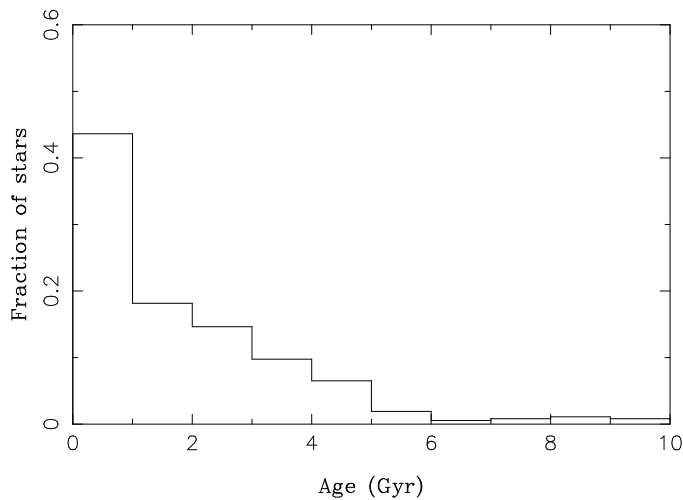}
   \caption{Age distribution of the central stars of planetary nebulae 
      for Method 5, based on the stellar velocities and on an relation 
      between the velocity dispersion and the stellar age.}
   \label{fig3}
   \end{figure}

\section{The PN samples}
\label{section3}

We have considered 4 different and independent samples of galacic planetary nebulae,
hereafter named as Sample A, B, C, and D. 

\subsection{Samples A, B, and C}

Sample A is the same sample used by Maciel et al. (\citeyear{mcu2003}, \citeyear{mlc2005}), 
which includes 234 well-observed nebulae in the solar neighbourhood and in the galactic 
disk. These objects have galactocentric distances in the range $4 <  R {\rm (kpc)} < 14$. 
The galactocentric distances $R$\ are calculated from the heliocentric distances and the
galactic coordinates in a straightforward way for a given value of the solar galactocentric
distance, which is usually adopted as  $R_0 = 8.0-8.5\,$kpc. The real uncertainty comes 
naturally from the heliocentric distances, so that we have adopted two different PN distance 
scales, as discussed below. Sample~B is a smaller sample containing 122 PN with sufficient data 
for Method~3 as used by Maciel et al. (\citeyear{mcu2003}). Sample~C is the largest of the two 
samples used by Maciel et al. (\citeyear{mrc2011}) for their Method~2, which corresponds to our 
Method~5 of the present paper. This sample contains all  nebulae with accurate radial velocities 
from the catalogue by Durand et al. (\citeyear{durand}). We have adopted the distances by 
Stanghellini et al. (\citeyear{ssv2008}) (SSV), which reduces the sample to 403 objects. Taking into 
account only those objects with accurate oxygen abundances, kinematic ages in the interval  
$0 < t ({\rm Gyr}) < 14$ and galactocentric distances in the range $3 < R ({\rm kpc})  < 12$, 
we have a final sample of 168 nebulae. The last restriction is important as the gradient apparently 
flattens out for $R < 3\,$kpc and probably also for $R > 12\,$kpc, so that we will be able to 
estimate the {\it average} disk gradient at any given epoch.

\subsection{Sample D}

Henry et al. (\citeyear{henry2010}) obtained a sample containing 124 planetary nebulae with 
homogeneously determined chemical abundances, which in principle allows the determination of 
more accurate gradients. These objects are located in the galactic disk, so that they are 
adequate to investigate the presence of radial abundance gradients. In their analysis, Henry 
et al. (\citeyear{henry2010}) derived an average gradient of $-0.058 \pm 0.006$ dex/kpc for 
the O/H ratio, accounting for uncertainties both in the oxygen abundances and in the radial 
distances, and using a detailed statistical procedure. The homogeneity of the sample derives
from the fact that all objects have been observed and analyzed by the same group, using the 
same observational and reduction techniques (see also Henry et al. \citeyear{henry2004} and 
Milingo et al. \citeyear{milingo2010}). The adopted distances come largely from the same source 
in the literature, namely the work by Cahn et al. (\citeyear{cks1992}), hereafter CKS, and the 
solar galactocentric distance was taken as 8.5 kpc. Considering the more recent SSV distance scale 
by Stanghellini et al. (\citeyear{ssv2008}), which includes data for 101 objects of the sample, 
Henry et al. (\citeyear{henry2010}) obtained a somewhat flatter gradient of $-0.042 \pm 0.004$ dex/kpc.

We have used the sample by Henry et al. (\citeyear{henry2010}) and calculated the individual 
ages using Methods 1, 3, and 5 above, using the oxygen abundances given in that paper. The 
N/O abundances needed for Method~3 were taken from the same group for consistency, as given 
in Henry et al. (\citeyear{henry2010}), Henry et al. (\citeyear{henry2004}), and Milingo et al.
(\citeyear{milingo2010}). For the objects not found in these samples we have used data
from our own compilations, as given by Maciel et al. (\citeyear{mci2010}) and
Quireza et al. (\citeyear{quireza2007}). Most of the objects in this sample have 
galactocentric distances $R < 15\,$kpc, which is particularly interesting, since very 
few anticentre nebulae with large galactocentric distances have been analyzed so far, 
and available results suggest a spatial variation of the gradients which is not fully 
understood  (see discussions by Henry et al. \citeyear{henry2010} and Costa et al. 
\citeyear{costa2004}).

\section{Results and discussion}
\label{section4}

\subsection{Estimating the time variation of the abundance gradients}

In this work we have  used the method initially proposed by Maciel et al. (\citeyear{mlc2005}), 
which consists in separating the planetary nebula samples into two groups according to their ages.
An age limit $t_L$ is defined and considered as a free parameter, adopting values 
in the range $1 \leq t_L \leq 14\,$Gyr. Then the  average O/H abundances and the magnitude of 
the radial gradients are estimated as functions of the age limit for each of the methods
1, 3, and 5. By comparing the average gradients of the \lq\lq younger\rq\rq\ group with those 
of the \lq\lq older\rq\rq\ group, we can in principle detect any systematic differences between 
the two groups considered. In practice the adopted range of the age limit is shorter than 
indicated above, since both for very young and very old objects one of the groups becomes 
very small, so that the results become statistically innacurate. However, as we will see in the 
following discussion, with age limits in the range $1 \leq t_L \leq 5\,$ Gyr in most cases 
the results show a well defined pattern, which is largely independent of the adopted age limit.
The formal uncertainties of the methods adopted here, as estimated by Maciel et al. 
(\citeyear{mci2010}, \citeyear{mrc2011}) are about 1--2 Gyr.

\subsection{Results for Samples A, B, C, and D}

The main results of this paper are shown in Tables 1--4 and in Figures 4--9. The tables show 
the complete results, as follows: Table 1 gives the results for the complete samples A, B, C, 
and D for Methods 1, 3, and 5. The selected sample is shown in column~1, the method used is given 
in column~2, the total number $N$ of objects considered is in column~3, the average O/H abundances 
and uncertainties in column~4, the derived oxygen gradient slope (dex/kpc) with uncertainties in 
column~5, and the correlation coefficient $r$ in column~6. The average abundances in column~4 are 
simple averages, calculated irrespective of the galactocentric distances. For sample~D we show 
the results using both the CKS and SSV distance scales.  Table~2 shows the results for samples A, 
B and C considering Methods 1, 3, and 5, referred to as M1, M3, and M5, respectively, taking into 
account two age groups, so that we have for each adopted age limit $t_L$ in Gyr (column~2) the 
number of objects  $N$ in the samples, the average O/H abundances, the derived gradients and 
correlation coefficients for both the  young (columns~3, 4, 5, and 6) and old (columns 7, 8, 9, 
and 10) groups. The corresponding results for Sample D are shown in Tables~3 and 4, for distances 
by CKS and SSV, respectively. In fact, the results of this paper concerning the average oxygen 
abundances and the O/H gradients are essentially the same for both distance scales by Cahn et al. 
(\citeyear{cks1992}) or Stanghellini et al. (\citeyear{ssv2008}). 

Examples of the procedure adopted here are given in Figures~4, 5, and 6 for Methods~1,
3, and 5, respectively. In these figures, we show  the average O/H abundances for the 
young and old groups as a function of the age limit for Sample D, using data by Henry et 
al. (2010) with distances by Cahn et al. (1992) (Figures~4a, 5a, and 6a). In Figures~4b, 
5b, and 6b we show the corresponding results for the oxygen gradient. In both cases the 
average uncertainty is shown on the bottom left. These figures are representative of 
all cases studied here, in the sense that there is generally a common pattern in the 
distribution of the average O/H abundances and the radial gradient with the age limit, 
which can be observed also in Tables 2, 3, and 4. 

Concerning the average abundances, in most cases the younger groups are either systematically 
more oxygen-rich than the older group or both groups have similar abundances, irrespective 
of the adopted age limit. The former situation is valid for Samples A, B, and C using Methods~1, 
3, and 5, respectively, as shown in Table~2. Also, for Sample~D, Tables~3 and 4 and Figure~4 
confirm this result for Method~1; for Method~3 the abundances of both groups are similar
within the uncertainties, and the reverse is observed for Method~5, although the differences 
are small. Based on theoretical grounds concerning the chemical evolution of the galactic disk, 
it is expected that the younger groups are more oxygen-rich, as shown in most cases considered 
here, in view of the age-metallicity relation observed in the galactic disk. However, the 
difference is usually small, frequently of the same order or even smaller than the average 
uncertainty, which can be seen for example in Figures~4 and 5. The inverse behaviour observed
for Method~5 is probably due to the fact that the ages estimated by this method are generally 
very low, as can be seen for example in Figure~3, which means that most objects tend to belong 
to the younger group, thus increasing the average abundances. It should also be mentioned that 
for the extreme values of the age limit in Tables~2, 3, and 4 one of the samples becomes 
underpopulated, so that the corresponding results are less reliable. In conclusion, the younger 
groups have generally slightly higher average abundances compared to the older groups, but 
the difference in most cases are small, which can be attributed to the well known dispersion 
in the age-metallicity relation in the Galaxy (Rocha-Pinto et al. \citeyear{rp2000}, 
\citeyear{rp2006}, Feltzing et al. \citeyear{feltzing}, Bensby et al. \citeyear{bensby}, and 
Marsakov et al. \citeyear{marsakov}).

Considering now the oxygen gradients shown in Tables~2, 3, and~4 and in the examples 
in Figures~4, 5,  and 6, we can observe that in most cases the gradients for both groups 
are similar within the uncertainties. This is always true for Methods~3 and 5 as 
applied to all corresponding samples, namely, Samples~B, C and D (CKS or SSV). 
Only for Method~1 we can notice that the older group seems to have steeper gradients 
compared to the younger groups both for Samples~A and D, and the difference may 
reach about 0.03 dex/kpc, as shown in Figure~4b. Here we may be observing the inverse 
behaviour of the average abundances discussed above, since Method~1 produces 
preferentially very large ages, so that most of the samples will contain relatively 
aged objects, which will increase the observed differences between the gradients.
A probably more important reason for the differences between the linear gradients
calculated by Method~1 is that this method assumes a relation involving the 
abundances, ages, and galactocentric distances (equation 2 of Maciel et al.
\citeyear{mcu2003}). Since the average abundances of both groups do not differ
appreciably, this relation implicitly assumes that the gradients of the older
groups are somewhat steeper than those of the younger groups, which is in fact
what is observed in Figure~4b, for example. In all other cases, as exemplified in 
Figures~5b and 6b, both gradients are similar. In fact, from the discussions on 
the age determinations given in the previous papers (Maciel et al. \citeyear{mci2010}, 
\citeyear{mrc2011}), we would expect Method~1 to be less reliable compared to Methods~3 
and 5. Method~5 is in principle more correct, since it is based on more robust 
correlations between the stellar ages and the kinematic properties, but the 
hypotheses made in Maciel et al. (\citeyear{mrc2011}) concerning the stellar 
proper motions probably lead to an overestimate of the number of very young objects, 
which indeed can be observed in Figure~3. Moreover, the age-velocity dispersion as 
proposed by the Geneva-Copenhagen survey (Holmberg et al. \citeyear{holmberg2009}) 
is less accurate for very young objects, so that we confirm that the results for 
Method~3, case~B, of Maciel et al. (\citeyear{mci2010}) are probably more accurate.
The differences between the observed gradients of the young and old groups are shown
in the examples of Figures~7--9, where we have considered Sample~D by Henry et al.
(\citeyear{henry2010}) with distances by CKS. In figures~7, 8, and 9 the adopted age
limits are 4.0, 2.5 and 2.5 Gyr, corresponding to Methods~1, 3, and 5, respectively.
The figures show the corresponding gradient (dex/kpc) and the correlation coefficient
in each case.

\section{Conclusions}
\label{section5}

Based on the results above, the main conclusions of this paper are: 

\noindent (i) The younger groups have similar or slightly higher oxygen abundances compared with the 
older groups, especially from the data of Table~2, where the differences may reach about 0.3 dex, 
higher than the average uncertainties. This can be explained by our current ideas on galactic 
chemical evolution, since the the observed differences are consistent with the dispersion in 
the age-metallicity relation, as discussed in the previous section.

\noindent (ii) The gradients of both groups are similar within the uncertainties, so that the radial 
gradient has not changed appreciably during the lifetimes of the objects considered in this paper, 
which extend to about 5 Gyr, approximately. Therefore, the PN gradient is not expected to be very 
different from the gradient observed in HII regions and cepheid variables, which seems to be 
supported by recent observational data on these objects (cf. Cescutti et al. \citeyear{cescutti2007}, 
Fu et al. \citeyear{fu2009}, and Pedicelli et al. \citeyear{pedicelli2009}). For example, 
model results by Cescutti et al. (\citeyear{cescutti2007}) indicate an O/H gradient of 
about $-0.035\,$dex/kpc for the galactocentric range considered here, which is similar to 
the compiled cepheid data and also to the present results, as can be seen for instance in 
Table~1. Similar results for cepheids and HII regions are also compiled by Fu et al. 
(\citeyear{fu2009}) and Colavitti et al. (\citeyear{colavitti2009}). The last reference
in particular presents some recent HII data displaying a  distribution very similar to the
cepheid data. Also, Pedicelli et al. (\citeyear{pedicelli2009}) present a detailed compilation
of Fe abundances in cepheids with an average slope of $-0.05\,$ dex/kpc. From our own recent
work on the oxygen to iron relation in the galactic disk (Maciel et al. \citeyear{mcr2013}),
we conclude that the oxygen gradient is approximately 20\% lower than the iron gradient, so that 
these results are also in agreement with our present results within the uncertainties.

Our main conclusion on the abundance gradients is in agreement with some recent work by Gibson et al. 
(\citeyear{gibson}) and Pilkington et al. (\citeyear{pilkington}) where it is shown on the basis
of different sets of observational data and theoretical models that the oxygen gradient 
apparently has remained approximately constant in the local universe, so that the magnitude 
of the gradients flattens out for redshift values close to zero.

It would be interesting to extend the present investigation to other elements observed in
planetary nebulae, such as Ne, Ar, and S. In fact, some preliminary results involving a more
restricted sample of nebulae for which a morphological classification is possible support
the present results, in the sense that no important differences are observed in the gradients
of these elements for younger and older objects. However, these results must still be viewed
with caution, as the samples are smaller and problems such as the \lq\lq sulphur anomaly\rq\rq
(cf. Henry et al. \citeyear{henry2004}) are still to be clarified.

It should be stressed that our goal in this paper is to investigate any {\it temporal
variations} of the gradient, and not to determine the actual magnitude of the
oxygen gradient. In fact, from the results shown in Tables~1-4, it is apparent
that the magnitude of the O/H gradient depends on the adopted sample, especially
considering that most PN samples are relatively small. However, it may be
concluded that the oxygen gradient is probably in the range $d{\rm (O/H)}/dR \simeq
-0.03$ to $-0.07$ dex/kpc, and our suggested average value is  $d{\rm (O/H)}/dR \simeq
-0.05$, which is essentially the gradient derived from the highly homogeneous
sample~D, as can be seen  from Tables~3 and 4.

\begin{table*}
\small
\caption[]{Results for the total samples.}
\label{table1}
\begin{flushleft}
\begin{tabular}{cccccc}
\noalign{\smallskip}
\hline\noalign{\smallskip}
SAMPLE & METHOD &  N & O/H & $d({\rm O/H})/dR$ & $r$ \\
\noalign{\smallskip}
\hline\noalign{\smallskip}
A       & 1  &   234   & $8.63\pm 0.26$ & $-0.04\pm 0.01$  & $-0.35$  \\
B       & 3  &   111   & $8.73\pm 0.20$ & $-0.08\pm 0.01$  & $-0.68$  \\
C       & 5  &   168   & $8.64\pm 0.24$ & $-0.03\pm 0.01$  & $-0.21$  \\
D (CKS) & 1  &   124   & $8.59\pm 0.21$ & $-0.04\pm 0.01$  & $-0.53$  \\
D (CKS) & 3  &   119   & $8.59\pm 0.22$ & $-0.04\pm 0.01$  & $-0.53$  \\
D (CKS) & 5  &   91    & $8.62\pm 0.20$ & $-0.04\pm 0.01$  & $-0.53$   \\
D (SSV) & 1  &   101   & $8.56\pm 0.22$ & $-0.03\pm 0.01$  & $-0.54$   \\
D (SSV) & 3  &   97    & $8.60\pm 0.22$ & $-0.04\pm 0.01$  & $-0.56$  \\
D (SSV) & 5  &   88    & $8.62\pm 0.20$ & $-0.03\pm 0.01$  & $-0.53$   \\
\noalign{\smallskip}
\hline
\end{tabular}
\end{flushleft}
\end{table*}

\begin{table*}
\small
\begin{changemargin}{-2cm}{-2cm}
\caption[]{Results for Samples A, B, C, Methods 1, 3, 5 (M1, M3, M5).}
\label{table2}
\begin{flushleft}
\begin{tabular}{cccccccccc}
\noalign{\smallskip}
\hline\noalign{\smallskip}
   &  &  &  & YOUNG GROUP & & & &  OLD GROUP & \\
SAMPLE & $t_L$ & N & O/H & $d({\rm O/H})/dR$ & $r$ & N & O/H & $d({\rm O/H})/dR$ & $r$\\
\noalign{\smallskip}
\hline\noalign{\smallskip}
A, M1 & 2.5  & 40& $8.65\pm 0.27$ & $+0.02\pm 0.02$  & $+0.13$ & 194 & $8.63\pm 0.26$ & $-0.06\pm 0.01$ & $-0.47$\\
   & 3.0  & 46   & $8.68\pm 0.28$ & $-0.00\pm 0.02$  & $-0.03$ & 188 & $8.62\pm 0.26$ & $-0.06\pm 0.01$ & $-0.52$\\
   & 3.5  & 64   & $8.74\pm 0.28$ & $-0.03\pm 0.01$  & $-0.26$ & 170 & $8.59\pm 0.24$ & $-0.07\pm 0.01$ & $-0.57$\\
   & 4.0  & 89   & $8.75\pm 0.25$ & $-0.04\pm 0.01$  & $-0.39$ & 145 & $8.56\pm 0.24$ & $-0.07\pm 0.01$ & $-0.63$\\
   & 4.5  & 137  & $8.73\pm 0.23$ & $-0.05\pm 0.01$  & $-0.49$ & 97  & $8.49\pm 0.23$ & $-0.08\pm 0.01$ & $-0.67$\\
   & 5.0  & 173  & $8.71\pm 0.23$ & $-0.05\pm 0.01$  & $-0.50$ & 61  & $8.43\pm 0.24$ & $-0.09\pm 0.01$ & $-0.74$\\
B, M3 & 3.5& 12  & $8.95\pm 0.23$ & $-0.09\pm 0.01$  & $-0.98$ & 99  & $8.70\pm 0.18$ & $-0.08\pm 0.01$ & $-0.73$\\
   & 4.0  & 34   & $8.87\pm 0.18$ & $-0.09\pm 0.01$  & $-0.90$ & 77  & $8.66\pm 0.17$ & $-0.08\pm 0.01$ & $-0.82$\\
   & 4.5  & 76   & $8.79\pm 0.19$ & $-0.09\pm 0.01$  & $-0.83$ & 35  & $8.59\pm 0.14$ & $-0.07\pm 0.01$ & $-0.84$\\
   & 5.0  & 95   & $8.75\pm 0.20$ & $-0.09\pm 0.01$  & $-0.80$ & 16  & $8.57\pm 0.11$ & $-0.07\pm 0.01$ & $-0.79$\\
C, M5 & 1.0 & 78 & $8.67\pm 0.23$ & $-0.02\pm 0.01$  & $-0.19$ & 90  & $8.62\pm 0.25$ & $-0.03\pm 0.01$ & $-0.22$\\
   & 1.5  & 91   & $8.67\pm 0.22$ & $-0.02\pm 0.01$  & $-0.15$ & 77  & $8.60\pm 0.25$ & $-0.03\pm 0.01$ & $-0.25$\\
   & 2.0  & 99   & $8.67\pm 0.22$ & $-0.02\pm 0.01$  & $-0.19$ & 69  & $8.60\pm 0.26$ & $-0.03\pm 0.01$ & $-0.23$\\
   & 2.5  & 112  & $8.66\pm 0.23$ & $-0.02\pm 0.01$  & $-0.18$ & 56  & $8.60\pm 0.25$ & $-0.03\pm 0.02$ & $-0.24$\\
   & 3.0  & 121  & $8.66\pm 0.23$ & $-0.02\pm 0.01$  & $-0.21$ & 47  & $8.59\pm 0.26$ & $-0.03\pm 0.02$ & $-0.19$\\
   & 3.5  & 127  & $8.66\pm 0.23$ & $-0.03\pm 0.01$  & $-0.22$ & 41  & $8.58\pm 0.26$ & $-0.02\pm 0.02$ & $-0.14$\\
   & 4.0  & 135  & $8.66\pm 0.23$ & $-0.03\pm 0.01$  & $-0.22$ & 33  & $8.58\pm 0.27$ & $-0.02\pm 0.02$ & $-0.17$\\
\noalign{\smallskip}
\hline
\end{tabular}
\end{flushleft}
\end{changemargin}
\end{table*}

\begin{table*}
\small
\begin{changemargin}{-2cm}{-2cm}
\caption[]{Results for Sample D, Data by Henry et al. (2010), CKS distances.}
\label{table3}
\begin{flushleft}
\begin{tabular}{cccccccccc}
\noalign{\smallskip}
\hline\noalign{\smallskip}
   &  &  &  & YOUNG GROUP & & &  & OLD GROUP & \\
 & $t_L$ & N & O/H & $d({\rm O/H})/dR$ & $r$ & N & O/H & $d({\rm O/H})/dR$ & $r$ \\
\noalign{\smallskip}
\hline\noalign{\smallskip}
M1 & 3.0  & 23   & $8.60\pm 0.21$ & $-0.07\pm 0.01$  & $-0.88$ & 101 & $8.58\pm 0.22$ & $-0.06\pm 0.01$ & $-0.66$\\
   & 3.5  & 45   & $8.64\pm 0.19$ & $-0.05\pm 0.01$  & $-0.82$& 79  & $8.56\pm 0.22$ & $-0.07\pm 0.01$ & $-0.77$\\
   & 4.0  & 78   & $8.63\pm 0.17$ & $-0.04\pm 0.01$  & $-0.70$& 46  & $8.51\pm 0.26$ & $-0.08\pm 0.01$ & $-0.88$\\
   & 4.5  & 97   & $8.63\pm 0.17$ & $-0.04\pm 0.01$  & $-0.66$& 27  & $8.43\pm 0.28$ & $-0.08\pm 0.01$ & $-0.94$\\
   & 5.0  & 111  & $8.60\pm 0.21$ & $-0.04\pm 0.01$  & $-0.64$& 13  & $8.44\pm 0.21$ & $-0.08\pm 0.01$ & $-0.91$\\
M3 & 1.0  & 60   & $8.53\pm 0.25$ & $-0.04\pm 0.01$  & $-0.59$& 59  & $8.64\pm 0.16$ & $-0.02\pm 0.01$ & $-0.43$\\
   & 1.5  & 84   & $8.58\pm 0.23$ & $-0.04\pm 0.01$  & $-0.55$& 35  & $8.60\pm 0.17$ & $-0.03\pm 0.01$ & $-0.47$\\
   & 2.0  & 96   & $8.58\pm 0.23$ & $-0.04\pm 0.01$  & $-0.55$& 23  & $8.60\pm 0.18$ & $-0.03\pm 0.02$ & $-0.39$\\
   & 2.5  & 99   & $8.58\pm 0.22$ & $-0.04\pm 0.01$  & $-0.54$& 20  & $8.61\pm 0.19$ & $-0.05\pm 0.02$ & $-0.52$\\
   & 3.0  & 102  & $8.58\pm 0.22$ & $-0.03\pm 0.01$  & $-0.52$& 17  & $8.62\pm 0.20$ & $-0.05\pm 0.02$ & $-0.62$\\
   & 3.5  & 105  & $8.58\pm 0.22$ & $-0.03\pm 0.01$  & $-0.52$& 14  & $8.60\pm 0.20$ & $-0.06\pm 0.02$ & $-0.68$\\
M5 & 1.0  & 44   & $8.59\pm 0.21$ & $-0.04\pm 0.01$  & $-0.54$& 47  & $8.64\pm 0.19$ & $-0.03\pm 0.01$ & $-0.49$\\
   & 1.5  & 53   & $8.60\pm 0.21$ & $-0.04\pm 0.01$  & $-0.54$& 38  & $8.64\pm 0.19$ & $-0.03\pm 0.01$ & $-0.49$\\
   & 2.0  & 64   & $8.61\pm 0.21$ & $-0.03\pm 0.01$  & $-0.50$& 27  & $8.64\pm 0.19$ & $-0.04\pm 0.01$ & $-0.60$\\
   & 2.5  & 73   & $8.61\pm 0.21$ & $-0.04\pm 0.01$  & $-0.53$& 18  & $8.65\pm 0.16$ & $-0.03\pm 0.02$ & $-0.45$\\
   & 3.0  & 75   & $8.61\pm 0.21$ & $-0.03\pm 0.01$  & $-0.52$& 16  & $8.65\pm 0.17$ & $-0.05\pm 0.02$ & $-0.60$\\
   & 3.5  & 78   & $8.61\pm 0.21$ & $-0.03\pm 0.01$  & $-0.52$& 13  & $8.65\pm 0.19$ & $-0.07\pm 0.02$ & $-0.64$\\
\noalign{\smallskip}
\hline
\end{tabular}
\end{flushleft}
\end{changemargin}
\end{table*}

\begin{table*}
\small
\begin{changemargin}{-2cm}{-2cm}
\caption[]{Results for Sample D, Data by Henry et al. (2010), SSV distances.}
\label{table4}
\begin{flushleft}
\begin{tabular}{cccccccccc}
\noalign{\smallskip}
\hline\noalign{\smallskip}
    &  &  &  & YOUNG GROUP & & &  & OLD GROUP & \\
 & $t_L$ & N & O/H & $d({\rm O/H})/dR$ & $r$ & N & O/H & $d({\rm O/H})/dR$ & $r$\\
\noalign{\smallskip}
\hline\noalign{\smallskip}
M1 & 3.0  & 30   & $8.54\pm 0.22$ & $-0.04\pm 0.01$  & $-0.74$ & 71  & $8.57\pm 0.23$ & $-0.05\pm 0.01$ & $-0.67$\\
   & 3.5  & 47   & $8.59\pm 0.20$ & $-0.04\pm 0.01$  & $-0.75$ & 54  & $8.54\pm 0.24$ & $-0.06\pm 0.01$ & $-0.79$\\
   & 4.0  & 68   & $8.59\pm 0.18$ & $-0.03\pm 0.01$  & $-0.69$ & 33  & $8.50\pm 0.29$ & $-0.07\pm 0.01$ & $-0.90$\\
   & 4.5  & 81   & $8.58\pm 0.18$ & $-0.03\pm 0.01$  & $-0.60$ & 20  & $8.49\pm 0.34$ & $-0.08\pm 0.01$ & $-0.93$\\
   & 5.0  & 91   & $8.57\pm 0.22$ & $-0.03\pm 0.01$  & $-0.61$ & 10  & $8.51\pm 0.25$ & $-0.09\pm 0.01$ & $-0.92$\\
M3 & 1.0  & 47   & $8.55\pm 0.27$ & $-0.04\pm 0.01$  & $-0.62$ & 50  & $8.65\pm 0.16$ & $-0.02\pm 0.01$ & $-0.41$\\
   & 1.5  & 68   & $8.60\pm 0.24$ & $-0.04\pm 0.01$  & $-0.57$ & 29  & $8.61\pm 0.18$ & $-0.03\pm 0.01$ & $-0.50$\\
   & 2.0  & 78   & $8.60\pm 0.23$ & $-0.04\pm 0.01$  & $-0.58$ & 19  & $8.61\pm 0.19$ & $-0.04\pm 0.02$ & $-0.46$\\
   & 2.5  & 80   & $8.60\pm 0.23$ & $-0.04\pm 0.01$  & $-0.57$ & 17  & $8.60\pm 0.20$ & $-0.05\pm 0.02$ & $-0.56$\\
   & 3.0  & 83   & $8.60\pm 0.23$ & $-0.04\pm 0.01$  & $-0.55$ & 14  & $8.62\pm 0.21$ & $-0.06\pm 0.02$ & $-0.67$\\
   & 3.5  & 85   & $8.60\pm 0.23$ & $-0.04\pm 0.01$  & $-0.55$ & 12  & $8.60\pm 0.22$ & $-0.06\pm 0.02$ & $-0.72$\\
M5 & 1.0  & 42   & $8.59\pm 0.21$ & $-0.04\pm 0.01$  & $-0.57$ & 46  & $8.63\pm 0.19$ & $-0.03\pm 0.01$ & $-0.45$\\
   & 1.5  & 51   & $8.60\pm 0.21$ & $-0.04\pm 0.01$  & $-0.56$ & 37  & $8.65\pm 0.18$ & $-0.03\pm 0.01$ & $-0.45$\\
   & 2.0  & 62   & $8.61\pm 0.21$ & $-0.04\pm 0.01$  & $-0.53$ & 26  & $8.66\pm 0.18$ & $-0.03\pm 0.01$ & $-0.56$\\
   & 2.5  & 71   & $8.61\pm 0.21$ & $-0.04\pm 0.01$  & $-0.55$ & 17  & $8.67\pm 0.14$ & $-0.02\pm 0.02$ & $-0.27$\\
   & 3.0  & 73   & $8.61\pm 0.21$ & $-0.03\pm 0.01$  & $-0.54$ & 15  & $8.67\pm 0.15$ & $-0.03\pm 0.02$ & $-0.38$\\
   & 3.5  & 76   & $8.61\pm 0.21$ & $-0.03\pm 0.01$  & $-0.54$ & 12  & $8.68\pm 0.16$ & $-0.05\pm 0.04$ & $-0.38$\\
\noalign{\smallskip}
\hline
\end{tabular}
\end{flushleft}
\end{changemargin}
\end{table*}

   \begin{figure}
   \centering
   \includegraphics[angle=-90, width=9.0cm]{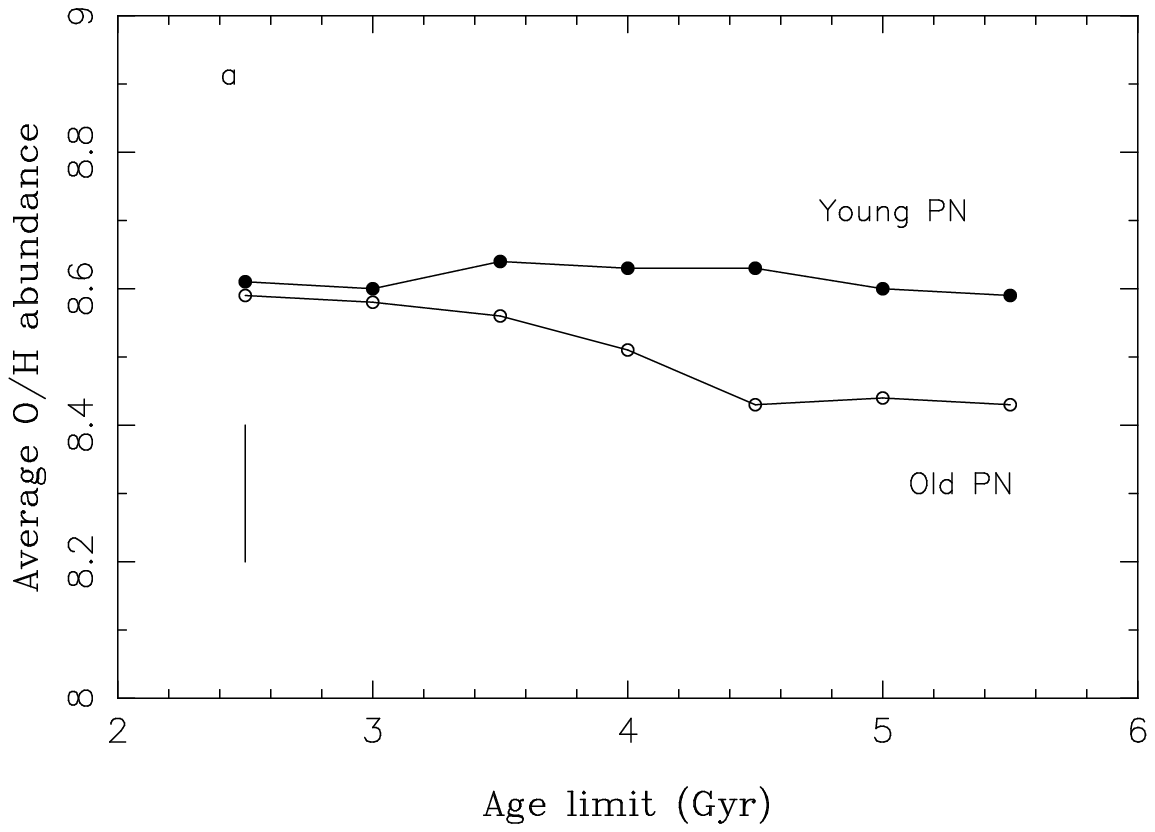}
   \includegraphics[angle=-90, width=9.0cm]{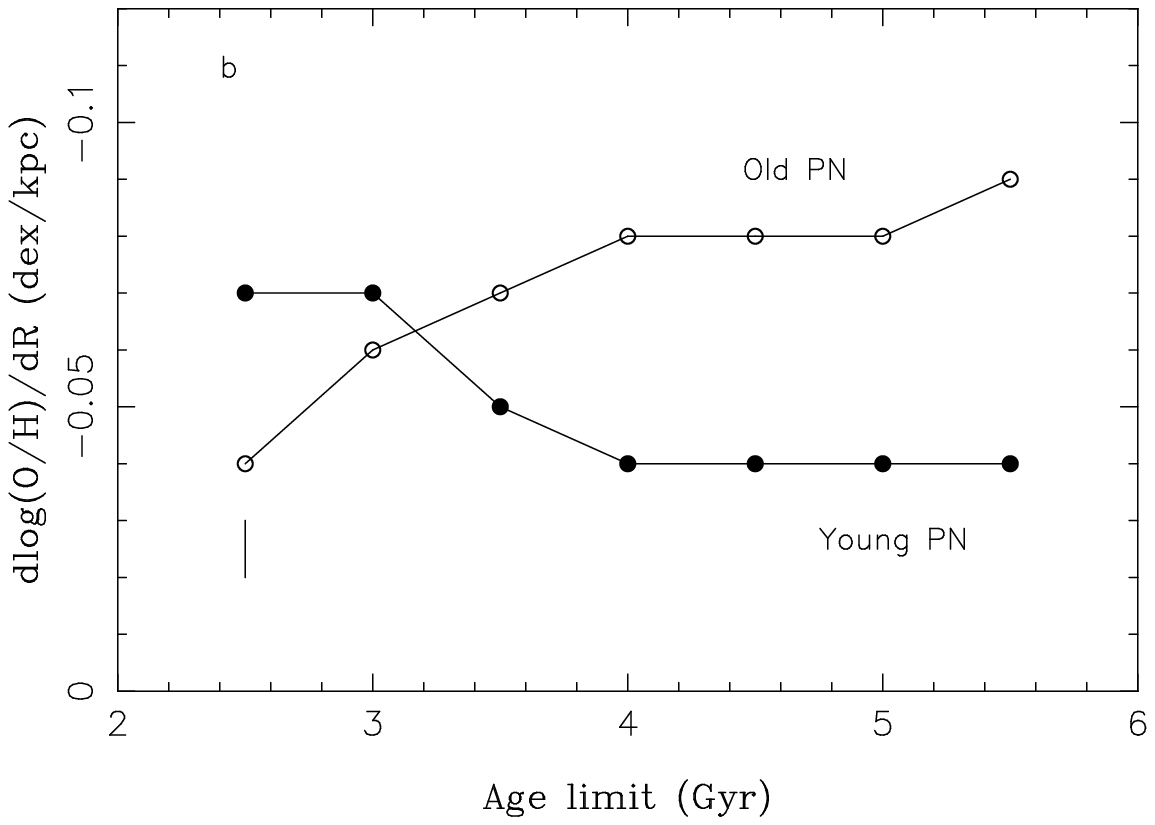}
   \caption{(a) Average O/H abundances for the young and old groups of PN as a function of 
   the age limit for Method 1, Sample D, data by Henry et al. (2010), and CKS distances). (b) 
   The same for the estimated O/H gradients (dex/kpc).}
   \label{fig4}
   \end{figure}

   \begin{figure}
   \centering
   \includegraphics[angle=-90, width=9.0cm]{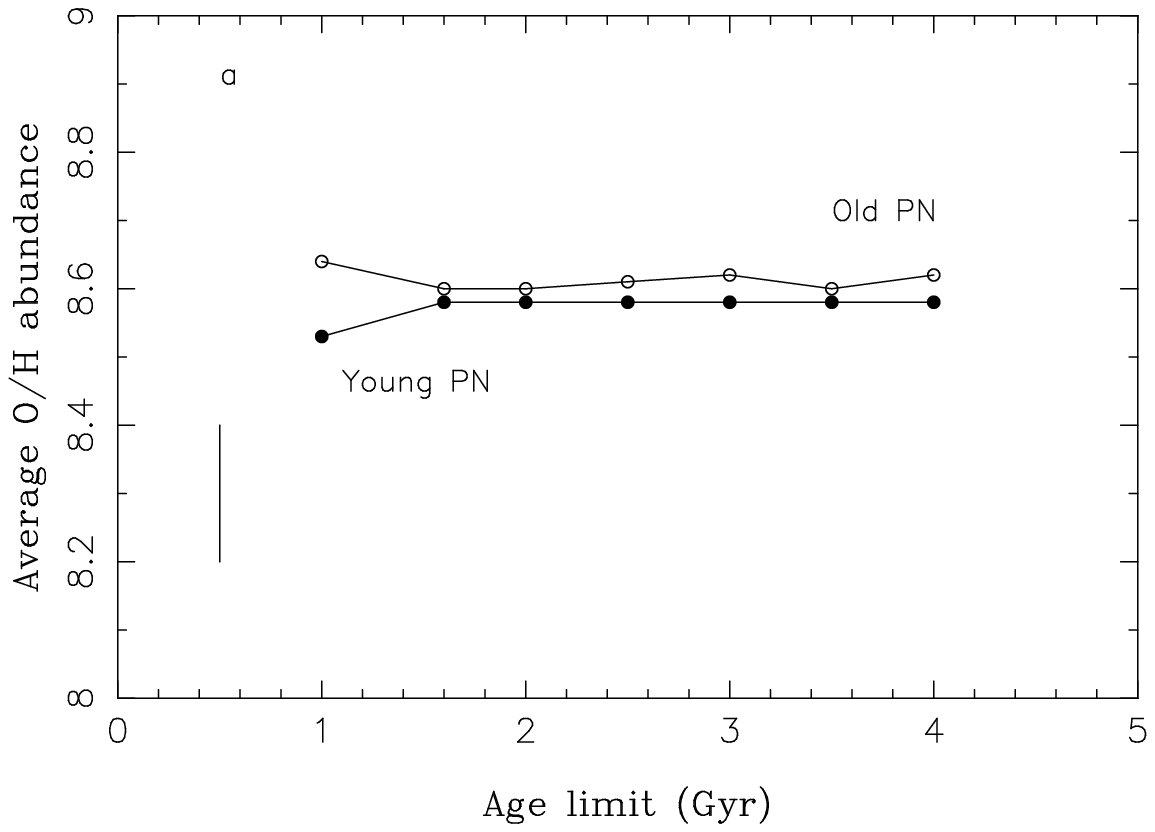}
   \includegraphics[angle=-90, width=9.0cm]{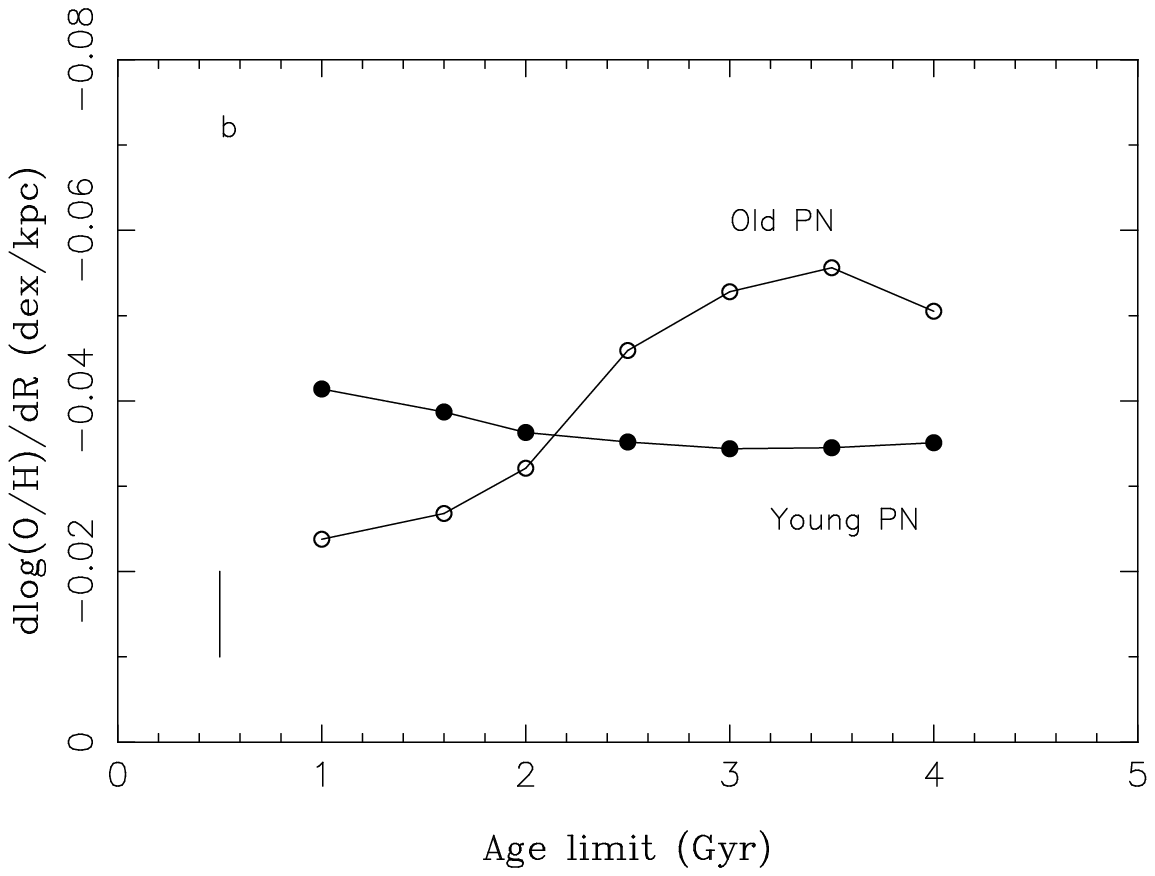}
   \caption{The same as Figure 4 for Method 3.}
   \label{fig5}
   \end{figure}

   \begin{figure}
   \centering
   \includegraphics[angle=-90, width=9.0cm]{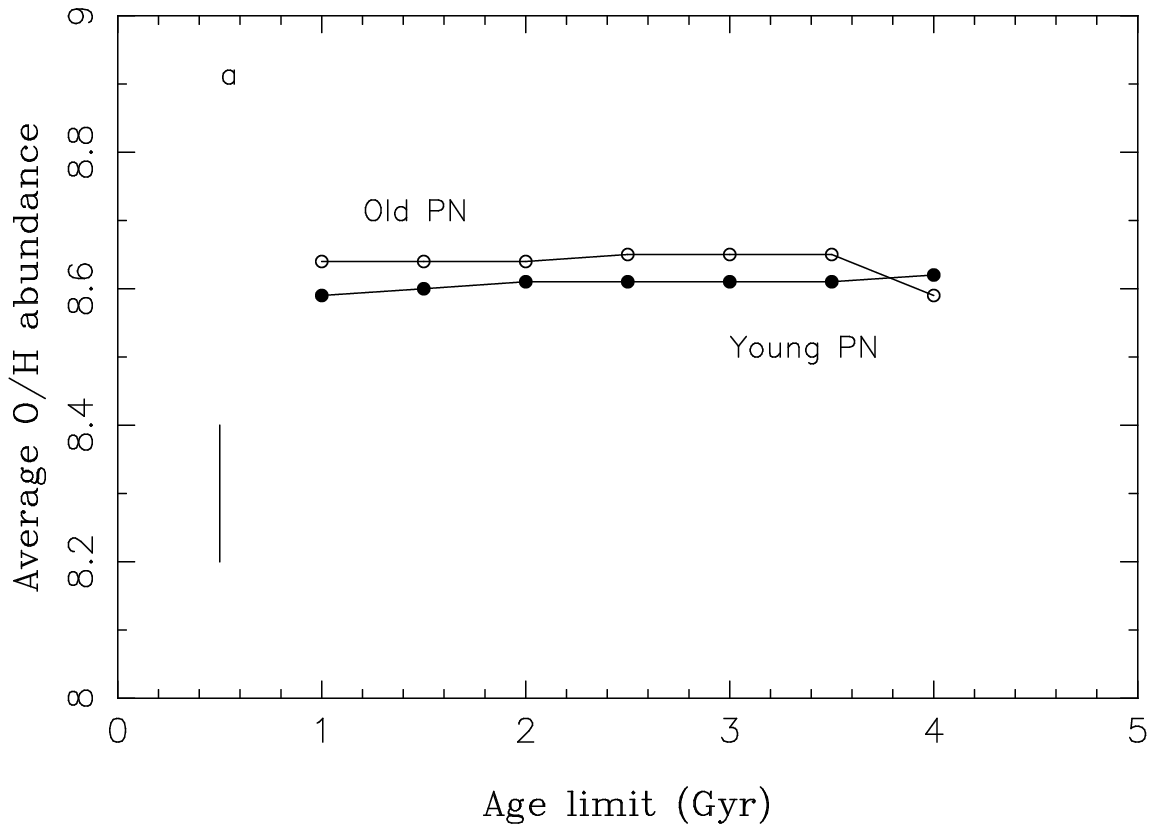}
   \includegraphics[angle=-90, width=9.0cm]{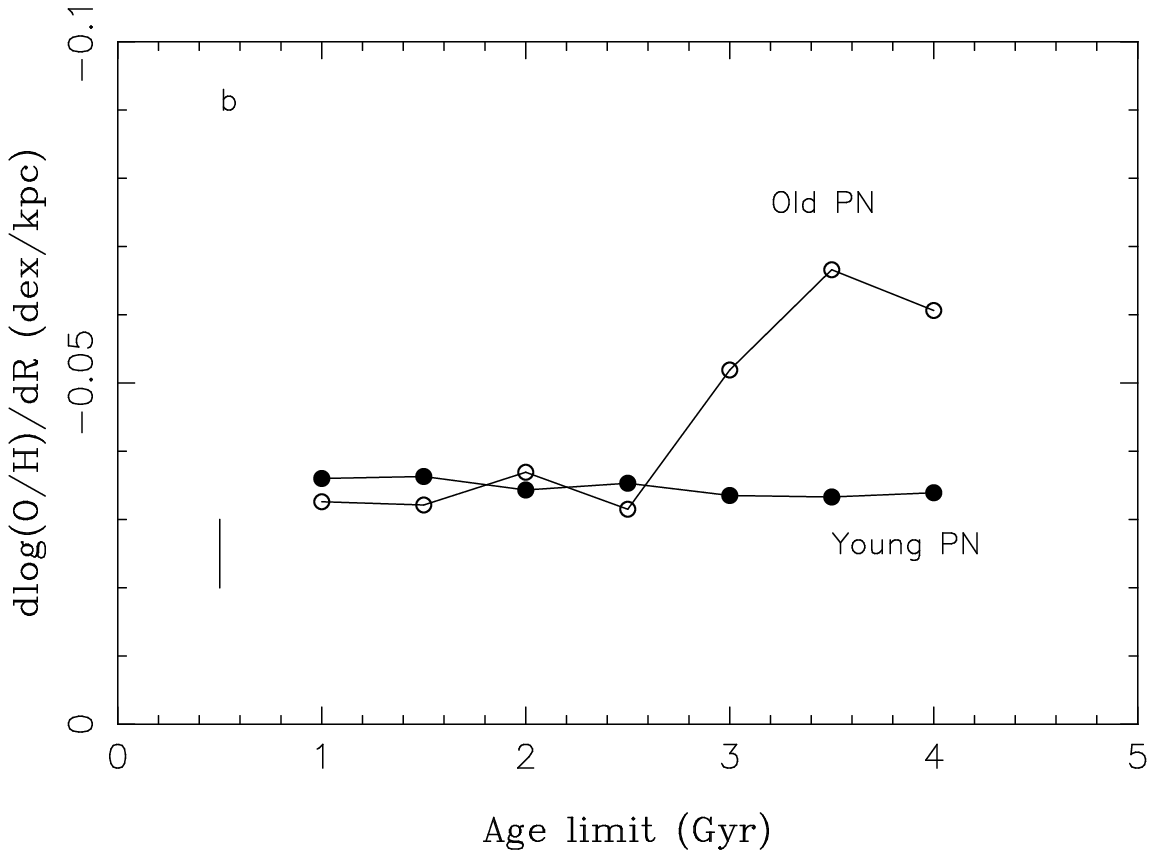}
   \caption{The same as Figure 4 for Method 5.}
   \label{fig6}
   \end{figure}

   \begin{figure}
   \centering
   \includegraphics[angle=-90, width=9.0cm]{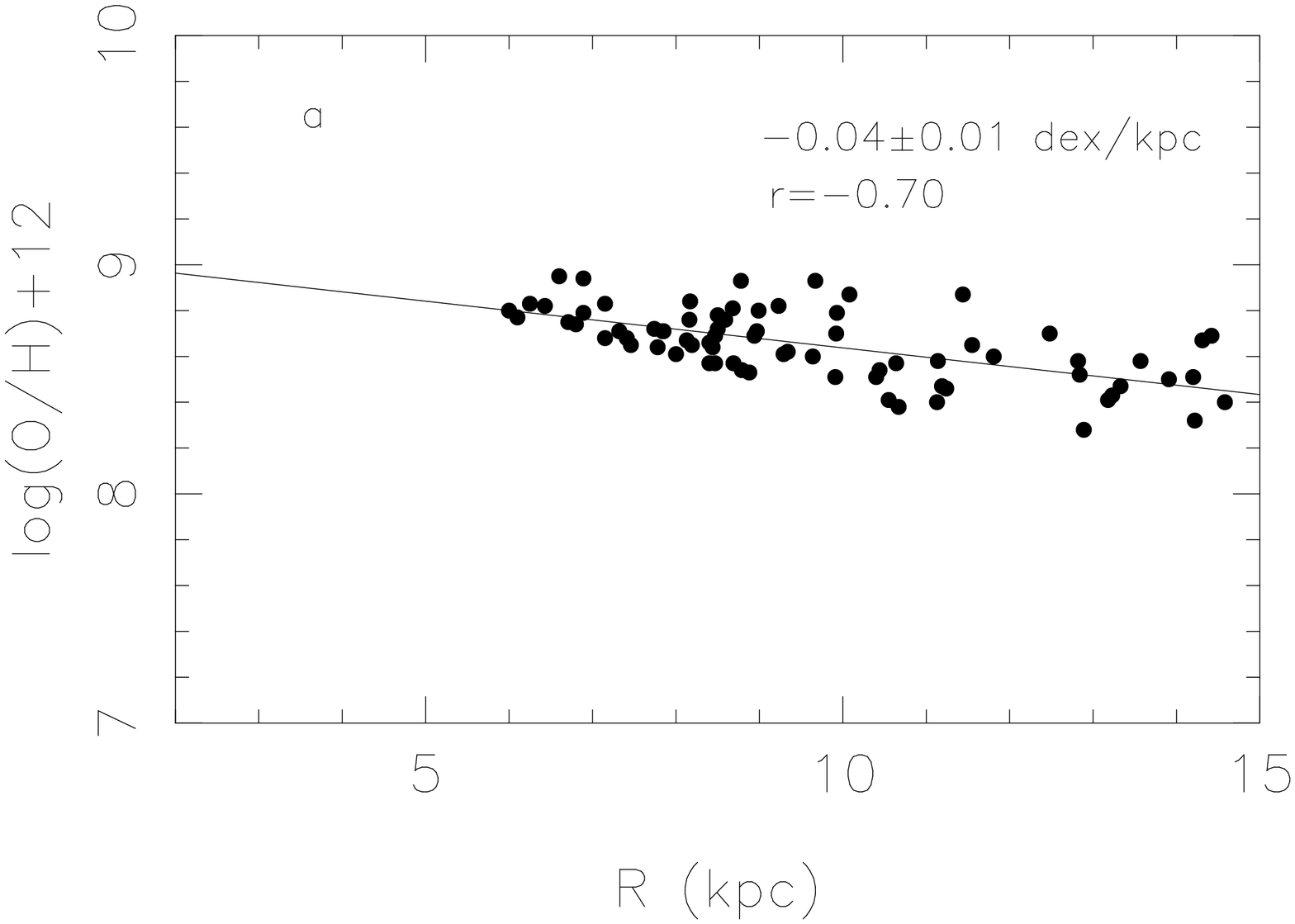}
   \includegraphics[angle=-90, width=9.0cm]{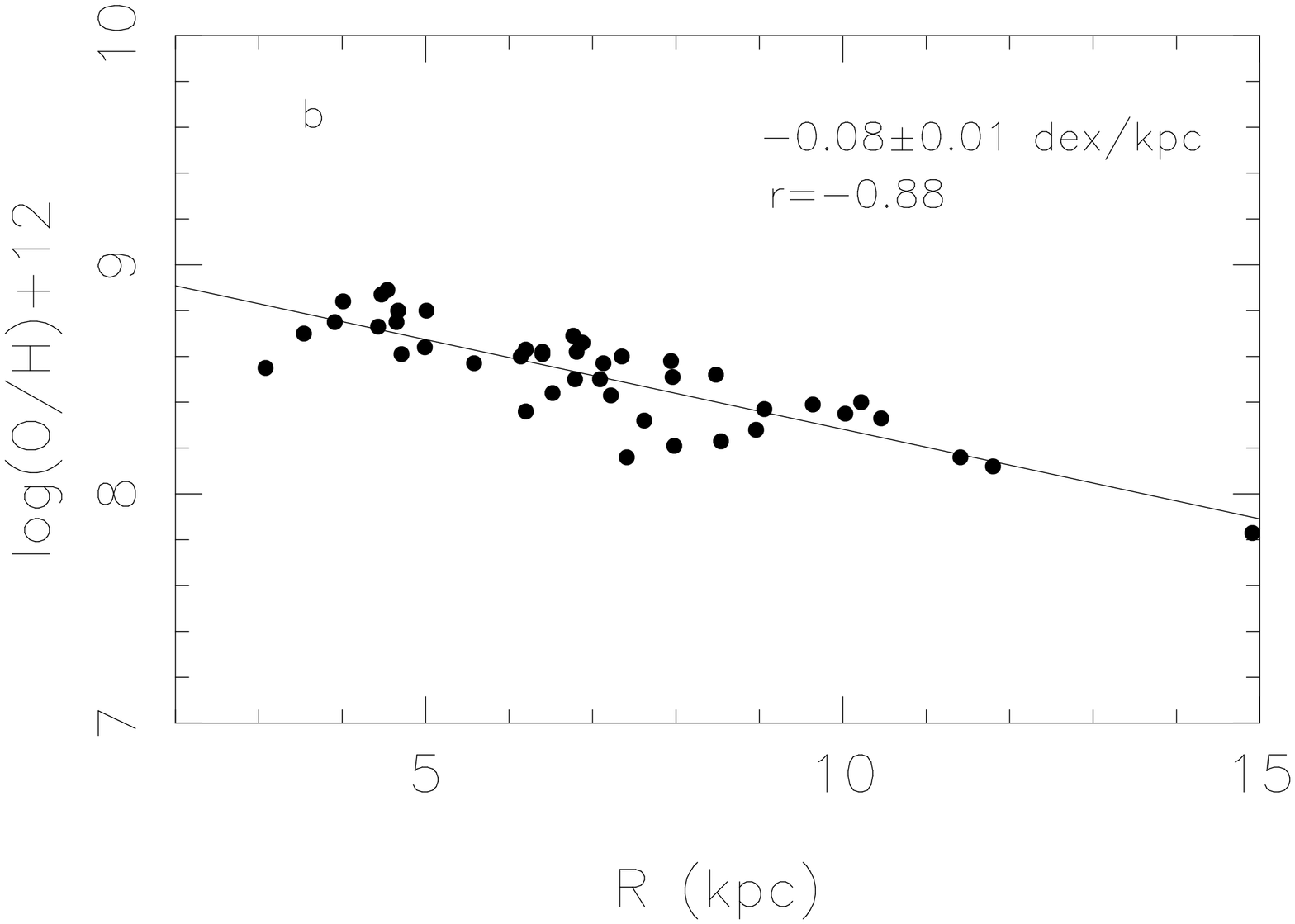}
   \caption{O/H abundances as a function of the galactocentric distance for Sample D, data 
   by Henry et al. (2010) and CKS distances for Method~1 and age limit 4.0 Gyr. The figures
   include the slopes (dex/kpc) and correlation coefficients $r$. (a) Young group, (b) Old group.}
   \label{fig7}
   \end{figure}

   \begin{figure}
   \centering
   \includegraphics[angle=-90, width=9.0cm]{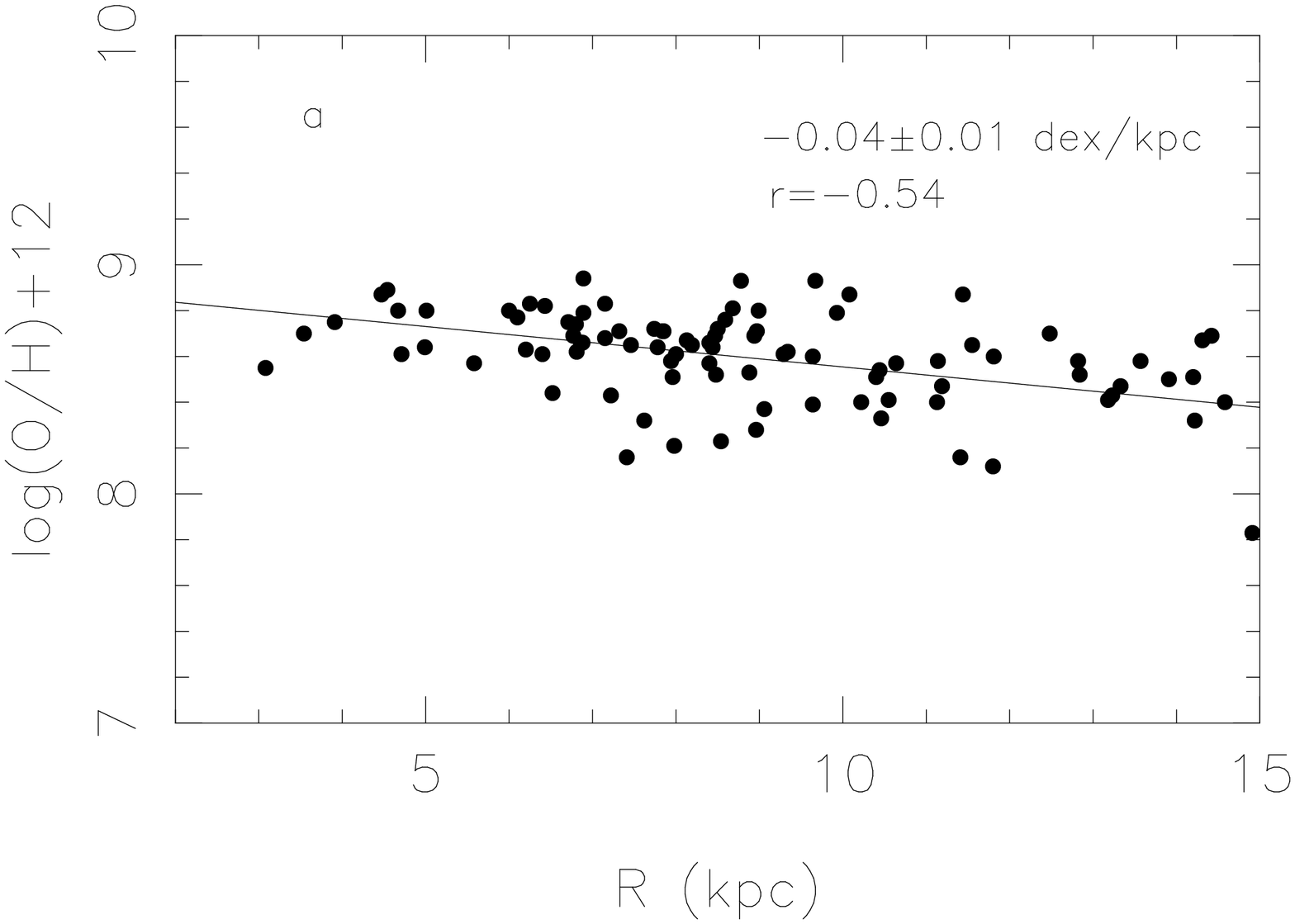}
   \includegraphics[angle=-90, width=9.0cm]{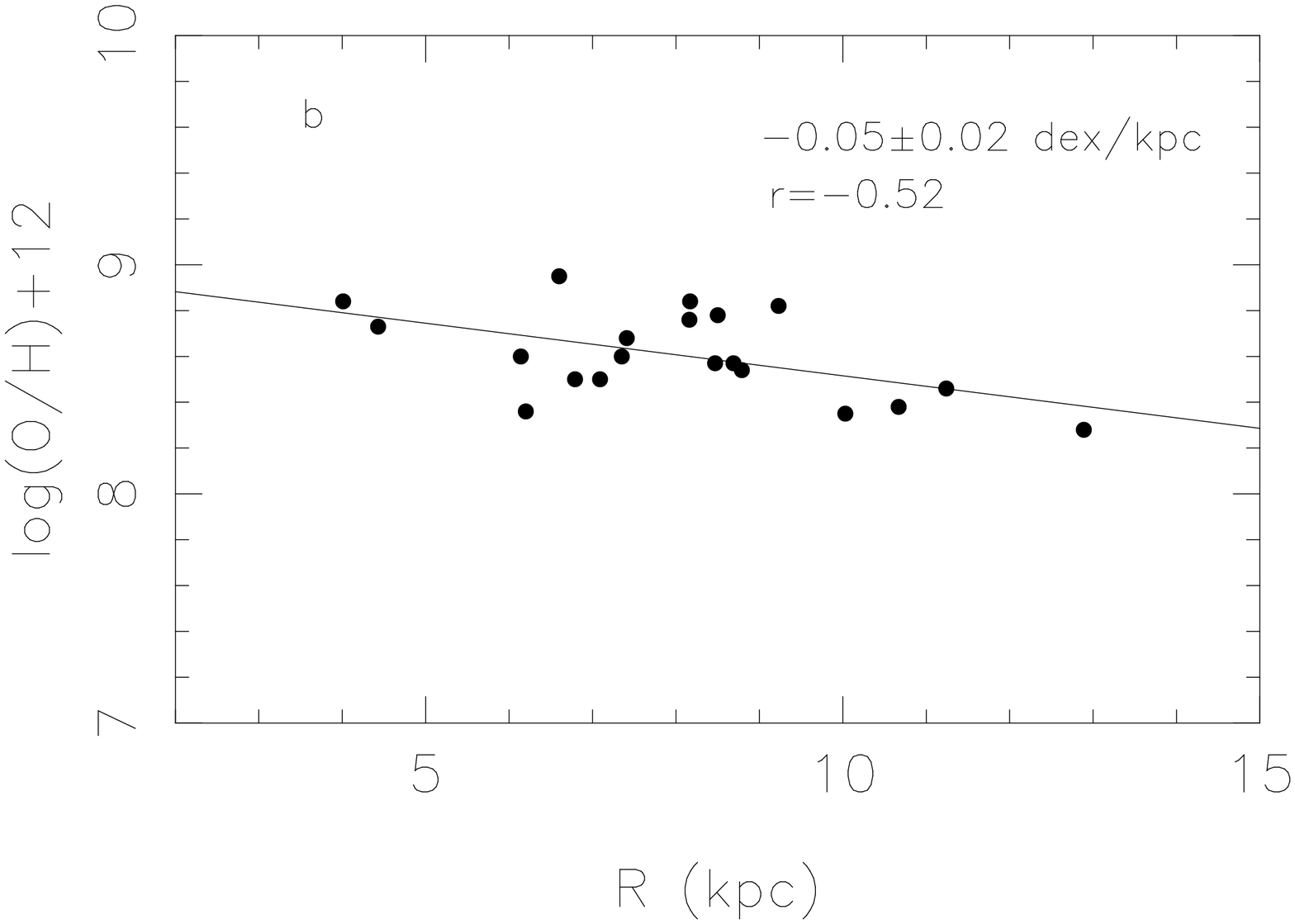}
   \caption{The same as Fig. 7 for Sample D, Method 3, age
    limit  2.5 Gyr.}
   \label{fig8}
   \end{figure}

   \begin{figure}
   \centering
   \includegraphics[angle=-90, width=9.0cm]{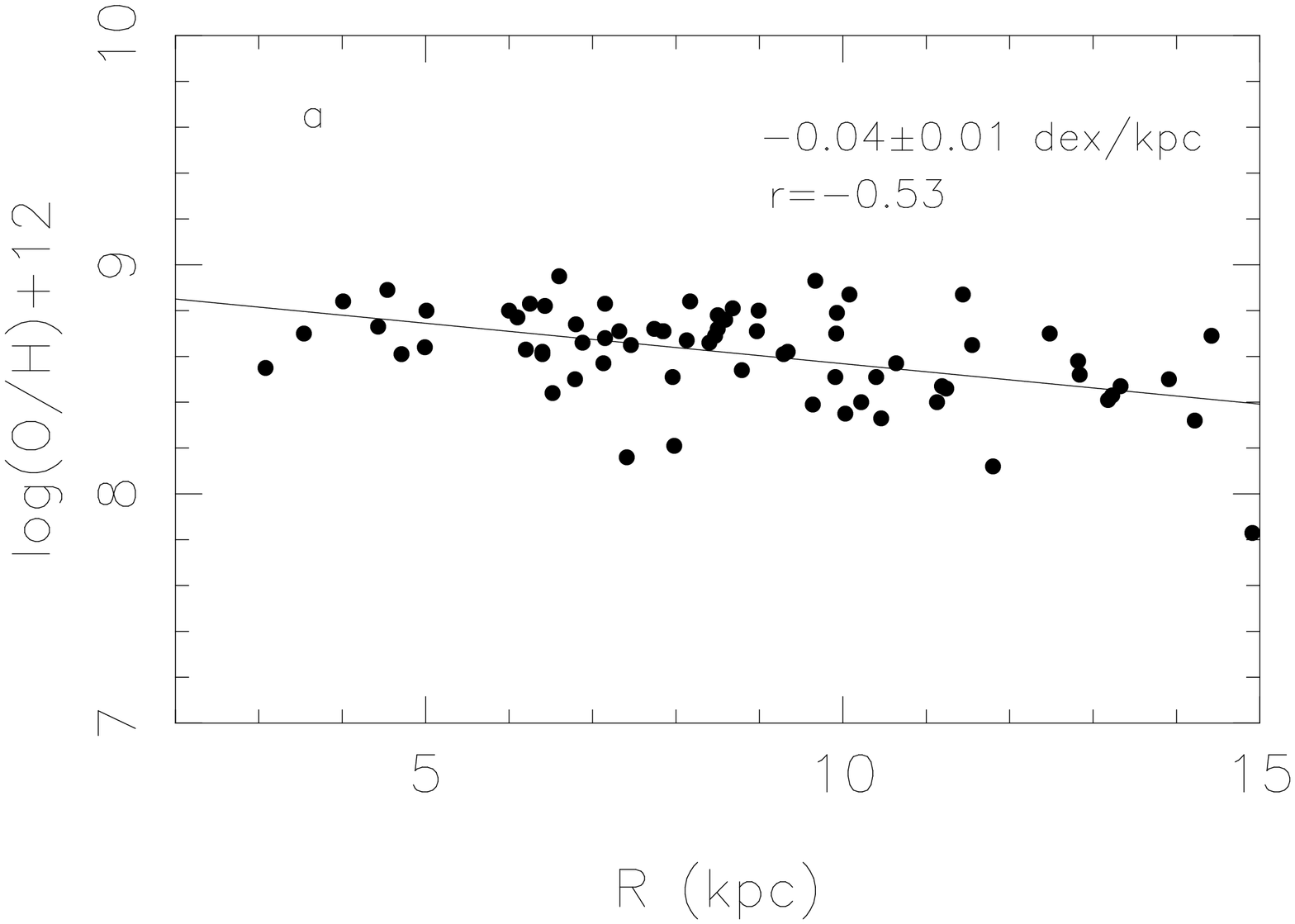}
   \includegraphics[angle=-90, width=9.0cm]{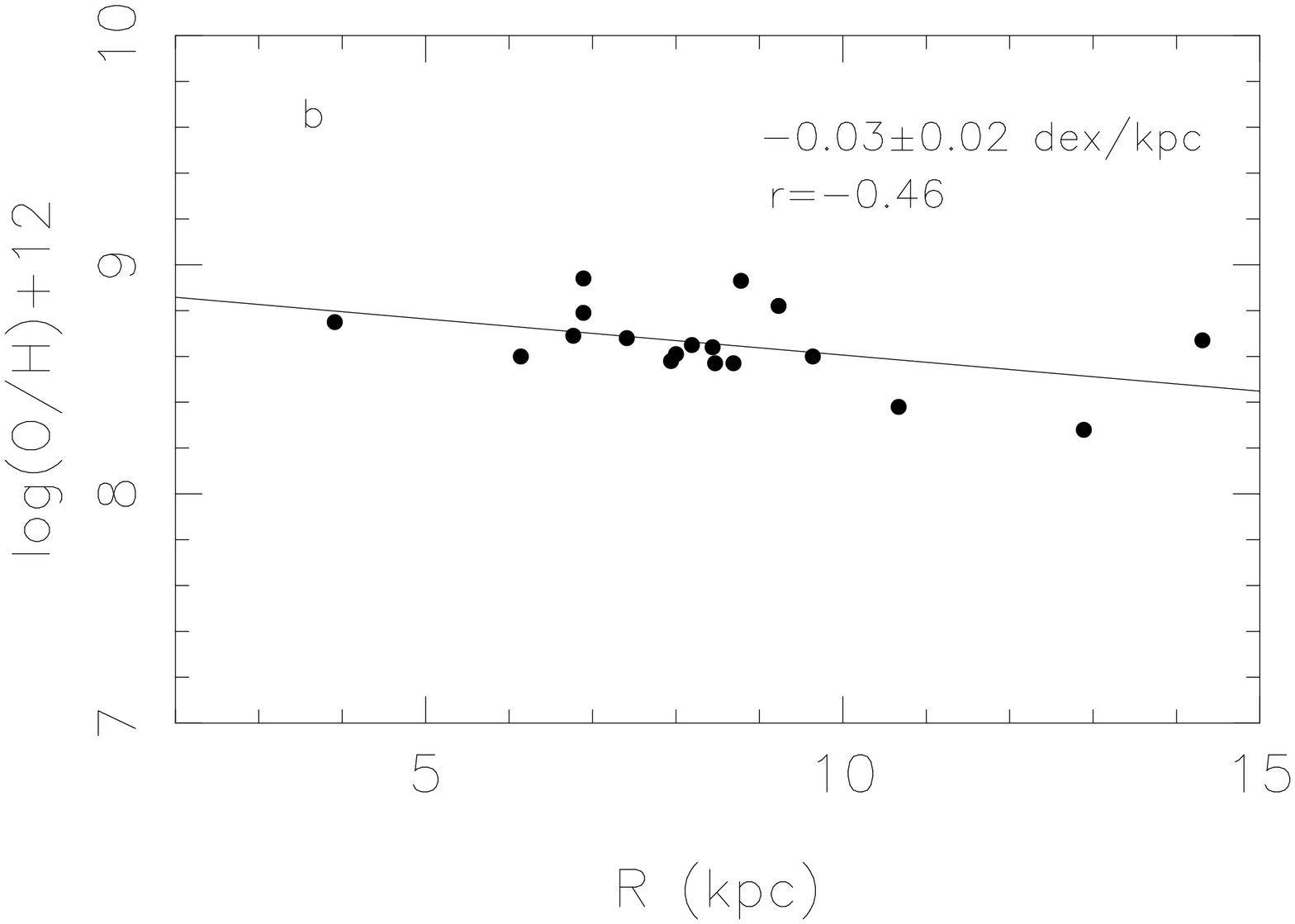}
   \caption{The same as Fig. 7 for Sample D, Method 5, age
    limit  2.5 Gyr.}
   \label{fig9}
   \end{figure}


\begin{thebibliography}


\bibitem[Andreuzzi et al. (2011)]{andreuzzi2011}
Andreuzzi, G., Bragaglia, A., Tosi, M. \& Marconi, G. 2011, MNRAS, 412, 1265

\bibitem[Andrievsky et al. (2013)]{andrievsky2013}
Andrievsky, S. M., L\'epine, J. R. D., Korotin, S. A., Luck, R. E., Kovtyukh, V. V., \& 
Maciel, W. J. 2013, MNRAS, 428, 3252

\bibitem[Bahcall \& Piran (1983)]{bahcall}
Bahcall, J. N., \& Piran, T. 1983, ApJ, 267, L77

\bibitem[Bensby et al. (2004)]{bensby}
Bensby, T., Feltzing, S., \& Lundstr\"om, I. 2004, A\&A, 421, 969

\bibitem[Cahn et al. (1992)]{cks1992}
Cahn, J. H., Kaler, J. B., \& Stanghellini, L. 1992, A\&AS, 94, 399 (CKS)

\bibitem[Cazetta and Maciel (2000)]{cazetta2000}
Cazetta, J. O., \& Maciel, W. J. 2000, Rev. Mex. Astron. Astrofis., 36, 3

\bibitem[Cescutti et al. (2007)]{cescutti2007}
Cescutti, G., Matteucci, F., Fran\c cois, P. \& Chiappini, C. 2007, A\&A, 462, 943

\bibitem[Chiappini et al. (2001)]{chiappini2001}
Chiappini, C., Matteucci, F., \& Romano, D. 2001, ApJ, 554, 1044

\bibitem[Colavitti et al. (2009)]{colavitti2009}
Colavitti, E., Cescutti, G., Matteucci, F., \& Murante, G. 2009, A\&A, 496, 429

\bibitem[Costa et al. (2004)]{costa2004}
Costa, R. D. D., Uchida, M. M. M., \& Maciel, W. J. 2004, A\&A, 423, 199

\bibitem[Dias et al. (2002)]{wilton}
Dias, W. S., Alessi, B. S., Moitinho, A., L\'epine, J. R. D. 2002, A\&A, 389, 871

\bibitem[Durand et al. (1998)]{durand}
Durand, S., Acker, A., \& Zijlstra, A., 1998, A\&S, 132, 13

\bibitem[Edvardsson et al. (1993)]{edvardsson}
Edvardsson, B., Andersen, J., Gustafsson, B., Lambert, D. L., Nissen, 
P. E., \& Tomkin, J. 1993, A\&A, 275, 101

\bibitem[Feltzing et al. (2001)]{feltzing}
Feltzing, S., Holmberg, J., \& Hurley, J. R. 2001, A\&A 377, 911

\bibitem[Fu et al. (2009)]{fu2009}
Fu, J., Hou, J. L., Yin, J., \& Chang, R. X. 2009, ApJ, 696, 668

\bibitem[Gibson et al. (2013)]{gibson}
Gibson, B. K., Pilkington, K., Bailin, J., Brook C. B., \& Stinson, G. S.
2013, Proceedings of Science, in press

\bibitem[Henry et al. (2004)]{henry2004}
Henry, R. B. C., Kwitter, K. B., \& Balick, B. 2004, AJ, 127, 2284

\bibitem[Henry et al. (2010)]{henry2010}
Henry, R. B. C., Kwitter, K. B., Jaskot, A. E., Balick, B., \&
Morrison, M. A. 2010, ApJ, 724, 748

\bibitem[Henry \& Worthey (1999)]{henry1999}
Henry, R. B. C., \& Worthey, G. 1999, PASP, 111, 919

\bibitem[Holmberg et al. (2007)]{holmberg2007}
Holmberg, J., Nordstr\"om, B., \& Andersen, J. 2007, A\&A,  475, 519

\bibitem[Holmberg et al. (2009)]{holmberg2009}
Holmberg, J., Nordstr\"om, B., \& Andersen, J. 2009, A\&A, 501, 941

\bibitem[Hou et al. (2000)]{hou2000}
Hou, J. L., Prantzos, N., \& Boissier, S. 2000, A\&A, 362, 921

\bibitem[Maciel \& Costa (2010)]{mc2010}
Maciel, W. J., \& Costa, R. D. D. 2010, IAU Symp. 265, Ed. K. Cunha, M. Spite, 
B. Barbuy (Cambridge: Cambridge University Press), 317 

\bibitem[Maciel et al. (2010)]{mci2010}
Maciel, W. J., Costa, R. D. D., \& Idiart, T. E. P. 2010, A\&A, 512, A19

\bibitem[Maciel et al. (2013)]{mcr2013}
Maciel, W. J., Costa, R. D. D., \& Rodrigues, T. S. 2013, ESO Workshop, The
Deaths of Stars and the Lives of Galaxies, 

http://www.eso.org/sci/meetings/2013/dslg2013/

\bibitem[Maciel et al. (2003)]{mcu2003}
Maciel, W. J., Costa, R. D. D., \& Uchida, M. M. M. 2003, A\&A, 397, 667

\bibitem[Maciel et al. (2005)]{mlc2005}
Maciel, W. J., Lago, L. G., \& Costa, R. D. D. 2005, A\&A, 433, 127

\bibitem[Maciel et al. (2011)]{mrc2011}
Maciel, W. J., Rodrigues, T. S., \& Costa, R. D. D. 2011, Rev. Mex. A\&A, 47, 401

\bibitem[Maciel et al. (2012)]{mrc2012}
Maciel, W. J., Rodrigues, T. S., \& Costa, R. D. D. 2012, IAU Symposium No. 283, 
Ed. A. Manchado, L. Stanghellini, D. Sch\"onberner (Cambridge: Cambridge University 
Press), 424 

\bibitem[Marsakov et al. (2011)]{marsakov}
Marsakov, V. A., Koval, V. V., Borkova, T. V., \& Shapovalov, M. V. 2011, Astron. 
Reports, 55, 667

\bibitem[Milingo et al. (2010)]{milingo2010}
Milingo, J. B., Kwitter, K. B., Henry, R. B. C., \& Souza, S. P. 2010, ApJ, 711, 619

\bibitem[Nordstr\"om et al.(2004)]{nordstrom}
Nordstr\"om, B., Mayor, M., Andersen, J. et al. 2004, A\&A, 418, 989

\bibitem[Pedicelli et al. (2009)]{pedicelli2009}
Pedicelli, S., Bono, G. Lemasle, B., Fran\c cois, P., Groenewegen, M. et al. 2009,
A\&A, 504, 81

\bibitem[Peimbert (1978)]{peimbert1978}
Peimbert, M., 1978, IAU Symp. 76, Ed. Y. Terzian (Dordrecht: Reidel), 215

\bibitem[Pilkington et al. (2012)]{pilkington}
Pilkington, K., Few, C. G., Gibson, B. K., Calura, F. et al. 2012,
A\&A, 540, A56

\bibitem[Quireza et al. (2007)]{quireza2007}
Quireza, C., Rocha-Pinto, H. J., \& Maciel, W. J. 2007, A\&A, 475, 217

\bibitem[Rocha-Pinto et al. (2000)]{rp2000}
Rocha-Pinto, H. J., Maciel, W. J., Scalo, J., \& Flynn, C. 2000, A\&A, 358, 850

\bibitem[Rocha-Pinto et al. (2006)]{rp2006}
Rocha-Pinto, H. J., Rangel, R. H. O., Porto de Mello, G. F., Bragan\c ca, G. A., 
\& Maciel, W. J. 2006, A\&A, 453, L9

\bibitem[Soderblom (2009)]{soderblom2009}
Soderblom, D. R. 2009, IAU Symp. 258, Ed. E. Mamajek, D. R. Soderblom, 
R. Wyse (Cambridge: Cambridge University Press), 1

\bibitem[Soderblom (2010)]{soderblom2010}
Soderblom, D. R. 2010, ARA\&A, 48, 581

\bibitem[Stanghellini et al. (2008)]{ssv2008}
Stanghellini, L., Shaw, R. A., \& Villaver, E. 2008, ApJ, 689, 194 (SSV)

\bibitem[Stasinska (2004)]{stasinska2004}
Stasi\'nska, G. 2004, Cosmochemistry: The melting pot of the elements, ed.
C. Esteban, R. J. García L\'opez, A. Herrero, F. S\'anchez (Cambridge:
Cambridge University Press), 115

\end{thebibliography}
\end{document}